\documentclass[letter,twocolumn]{jpsj3}
\usepackage{txfonts}
\usepackage[usenames]{color}

\title{Finite-Temperature Crossover Phenomenon \\ 
in the $S=1/2$ Antiferromagnetic Heisenberg Model on the Kagome Lattice}

\author{Tokuro Shimokawa\thanks{t.shimokaw@gmail.com} and Hikaru Kawamura}
\inst{Department of Earth and Space Science, Graduate School of Science, Osaka University,\\
 Toyonaka, Osaka 560-0043, Japan} 

\abst{Thermal properties of the $S$=1/2 kagome Heisenberg antiferromagnet at low temperatures are investigated by means of the Hams-de Raedt method for clusters of up to 36 sites possessing a full symmetry of the lattice. The specific heat exhibits, in addition to the double peaks, the third and the fourth peaks at lower temperatures. With decreasing the temperature, the type of the magnetic short-range order (SRO) changes around the third-peak temperature from the $\sqrt{3} \times \sqrt{3}$ to the $q=0$ states, suggesting that the third peak of the specific heat is associated with a crossover phenomenon between the spin-liquid states with distinct magnetic SRO. Experimental implications are discussed.}

\begin{document}
\maketitle

Geometrically frustrated magnets have attracted special interest due to its unique and novel ordering properties. Among them, kagome antiferromagnets have long been studied extensively. Especially, much recent interest has been paid to the quantum spin-1/2 nearest-neighbor (n.n.) antiferromagnetic (AF) Heisenberg model on the kagome lattice because of the possible realization of a quantum spin-liquid (QSL) state having no magnetic long-range order. A large number of theoretical studies performed to understand the nature of its ground state have lead to various competing scenarios on the nature of its ground state, including the $\mathbb{Z}_2$ spin liquid~\cite{Sachdev,Jiang,Lu,Yan,Depenbrock,Jiang2,Nishimoto}, the algebraic U(1) spin liquid~\cite{Hastings,Ran,Nakano,Iqbal,Iqbal2}, the chiral spin liquid~\cite{Messio}, the valence bond crystal~\cite{Marston,Singh,Evenbly}, {\it etc\/}. The true situation, however, still remains unclear.

 Along with such intensive studies on the ground state, thermal properties of the $S=1/2$ kagome antiferromagnetic Heisenberg (KAH) model at finite temperatures have also attracted much attention. A highlight issue might be the exotic temperature ($T$) dependence of the specific heat, which exhibits multiple peaks. Namely, earlier numerical studies based on an exact diagonalization (ED) method \cite{Elser, Zeng} or a decoupled-cell Monte Carlo (MC) simulation~\cite{Zeng} indicated that, in addition to the broad peak at a higher $T$, the specific heat exhibited the second peak at a lower $T$. Whether this second peak identified for small-size systems really survives in the thermodynamic limit had been examined by various calculations: Mentioning some of them, the ED method up to 24 spins~\cite{Elser, Zeng, Elstner, Misguich, Sindzingre}, a high-$T$ expansion~\cite{Elstner, Misguich2}, a high-$T$ entropy method~\cite{Misguich}, an approximate effective-Hamiltonian method~\cite{Zeng2, Sindzingre2}, a transfer-matrix MC method~\cite{Nakamura} and a linked-cluster algorithm~\cite{Rigol1, Rigol2}.

\begin{figure}[t]
  \includegraphics[width=8 cm,angle=0]{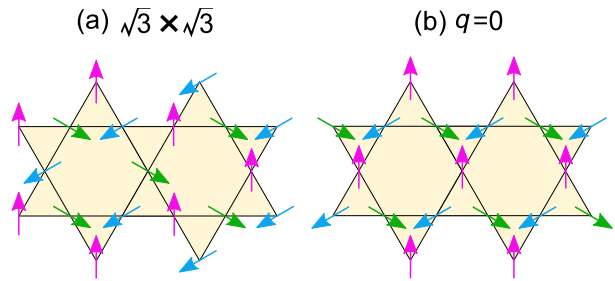}
 \caption{(Color online) (a) The $\sqrt{3} \times \sqrt{3}$, and (b) the $q=0$ spin structures.}
 \label{}
\end{figure}

 Recently, Sugiura and Shimizu have succeeded in computing the specific heat of the model up to the sizes $N=27$ and 30~\cite{Sugiura}, by using the imaginary-time version of the equation-of-motion method (the Hams-de Raedt algorithm), a powerful numerical technique of computing thermal properties of the quantum model at finite $T$ developed some time ago~\cite{Imada,Hams} (Sugiura and Shimizu called the method the canonical thermal pure quantum state method). It was then observed that, on increasing the system size up to $N=30$, the second peak was appreciably suppressed, with only a shoulder-like structure remaining~\cite{Sugiura}. Similar behavior was reported also by a finite-$T$ Lanczos method~\cite{Jaklic} applied to the $S=1/2$ KAH model of $N\leq 30$~\cite{Munehisa}.

 In the present Letter, we wish to investigate the finite-$T$ properties of the $S=1/2$ KAH model by means of the Hams-de Raedt method, paying attention not only to the multiple-peak problem of the specific heat, but also to the type of the magnetic SRO as mentioned below. We extend the cluster size up to 36 spins possessing a full symmetry of the lattice, exceeding the previous works. While the method could provide us exact information for frustrated quantum systems, special care might be taken to specific circumstances peculiar to the KAH model. For example, possible significance of the $\sqrt{3} \times \sqrt{3}$ [see Fig.~1(a)] SRO calls for finite-size clusters of multiples of nine, while the possible singlet ground state calls for even-$N$ clusters. In fact, the maximum sizes treated in recent exact finite-$T$ calculations, $N=24-30$, do not satisfy these requirements, and the results might be subject to stronger finite-size effects~\cite{Elser, Zeng, Elstner, Misguich, Sugiura}. The maximum size treated in the present work $N=36$ meets these criteria and is favorable in that respect. Indeed, the 36-spin cluster possesses the full symmetry of infinite kagome lattice under periodic boundary conditions.

We find that the second peak of the specific heat persists in the 36-spin cluster, even a bit more enhanced than that in the 27- and 30-spin clusters, suggesting that the second peak (or the shoulder) persists in the continuum limit. In addition, we observe in the 36-spin cluster the third and the fourth peaks at lower $T$~\cite{Sindzingre}. Interestingly, the third peak turns out to be associated with a finite-temperature crossover phenomenon between the two distinct magnetic SRO states, {\it i.e.\/}, the ones with the $\sqrt{3} \times \sqrt{3}$ SRO at higher $T$ and the $q=0$ [see Fig.~1(b)] SRO at lower $T$.

 Our model is the $S$=1/2 AF Heisenberg model on the kagome lattice, whose Hamiltonian is given by

\begin{eqnarray}
\mathcal{H}=J\sum_{i,j}{\bf S}_i \cdot {\bf S}_j
\end{eqnarray}
where ${\bf S}_i=(S_i^x,S_i^y,S_i^z)$ is a spin-1/2 operator at the $i$-th site on the lattice, and $J=1$ is the nearest-neighbor AF coupling. We treat several finite-size kagome clusters up to 36 spins with periodic boundary conditions. In computing the $T$ dependence of various physical quantities, we employ the Hams-de Raedt method~\cite{Imada,Hams,Sugiura}. This method allows us to compute physical quantities by treating a small number of quantum states instead of taking an ensemble average over a full spectrum of the Hilbert space. While in the most direct ED method the memory limitation prevents us from treating more than 20 spins even for $S$=1/2, the method enables us to treat systems containing about 40 spins.
\begin{figure}[t]
  \includegraphics[width=7.0cm,angle=0]{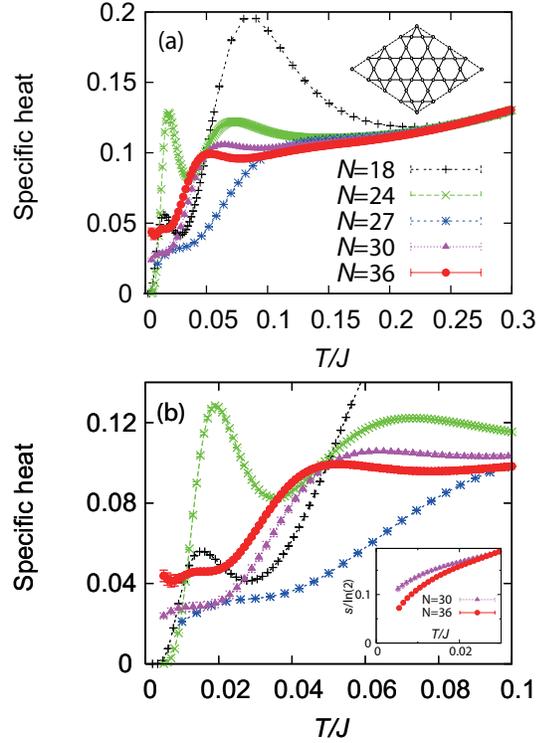}
 \caption{(Color online) The temperature dependence of the specific heat of $N=$18, 24, 27, 30 and 36 in the temperature regions of (a) $0 \leq T \leq 0.3$, and of (b) $0 \leq T \leq 0.1$. The 36-site cluster is shown in the inset of (a) \cite{shapes}. The lower-temperature dependence of the entropy per site for $N=30$ and 36 is shown as the inset of (b).
}
 \label{}
\end{figure}

 We briefly describe the computational method. A set of pure states for the inverse temperature  $\beta=1/T$ and the system size $N$, $|\beta, N \rangle $, is constructed by operating an operator on a set of initial random vectors as $|\beta, N \rangle = {\rm exp}[-\beta \mathcal{H} /2] |\psi_0 \rangle$, where the initial vectors are given by $|\psi_0 \rangle=\sum_{i=1}^{2^N} c_i |i \rangle$ with \{$c_i$\} being random complex numbers satisfying the normalization condition $\sum_i|c_i|^2=1$ and with $\{|i\rangle \}$ being an arbitrary orthonormal basis set of the Hilbert space of $\mathcal{H}$. Hams and de Raedt proved that the standard thermal average of a physical quantity $\hat{A}$ was given by 
\begin{eqnarray}
\langle \hat{A} \rangle_{\beta,N}= \overline{\langle \beta, N| \hat{A} | \beta, N \rangle} / \overline{\langle \beta, N|\beta, N \rangle},
\end{eqnarray}
where the overline denotes the average over the initial random vectors~\cite{Hams}. When this random average is performed over finite number $I$ of realizations of initial vectors, as is necessarily the case in real numerical calculations,  the deviation from the true value decays as $\sim 1/\sqrt{ID}$ where $D$ represents the dimension of the entire Hilbert space of the model, $D=2^N$ in the present case~\cite{Hams}. This relation means that, for larger system size $N$, fairly accurate value can be obtained even from smaller-$I$ calculations. In our computation, the average over initial random vectors is taken over 200 ($N$=18), 40 ($N$=24), 20 ($N$=27), 10 ($N$=30) and 3 ($N$=36) realizations.

\begin{figure}[t]
 \includegraphics[width=6.8cm, angle=0]{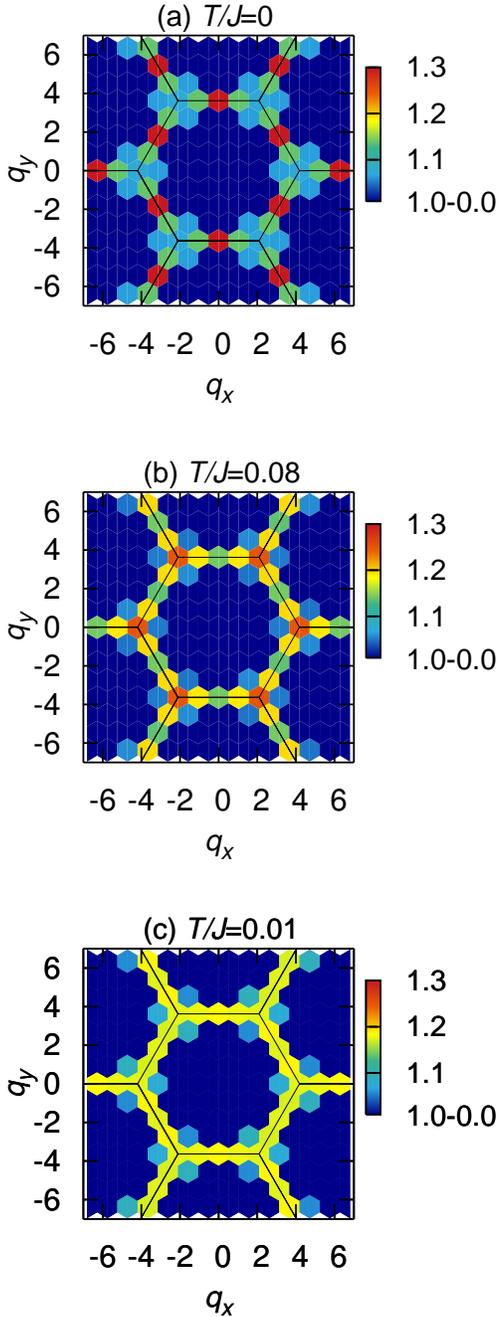}
 \caption{(Color online) Intensity plots of the static spin structure factor $S({\bf q}, \beta)$ in the wave-vector ($q_x$,$q_y$) plane for $N=36$ at temperatures (a) $T=0$, (b) $T=0.08$, and (c) $T=0.01$. The $T=0$ data are obtained by the ED method. The length unit is taken to be the n.n. distance of the kagome lattice. The solid black line depicts the zone boundary of the extended BZ. Note that the color corresponding to the intensities less than one is all set to blue. 
}
\label{}
\end{figure}

 The computed specific heat is displayed in Fig.~2(a) in the temperature region $T\leq 0.3$ for the sizes $N=$18, 24, 27, 30 and 36, Fig.~2(b) being its low-$T$ ($T\leq 0.1$) magnification. Our results for $N=18$ and 24 agree with the ED results of Refs.~\cite{Elstner, Misguich, Sindzingre}, and those for $N$=27 and 30 with the results of Ref.~\cite{Sugiura}. 

 A broad first peak arises around $T \sim 0.7$ (not indicated in Figs.2), while the second peak arises at $T\simeq 0.05-0.1$, whose location gradually moves to lower $T$ on increasing $N$. 
Interestingly, on increasing $N$ for $18\leq N\leq 30$, the sharpness of the second peak tends to be gradually suppressed, while it is a bit more enhanced for $N=36$ than for $N=30$. This recovery of the second peak might be related to the fact that the $N=36$ cluster retains a full symmetry of the infinite kagome lattice allowing for the $\sqrt{3} \times \sqrt{3}$ structure accommodated. As shown below, we find that the magnetic SRO around the second-peak temperature is indeed the $\sqrt{3} \times \sqrt{3}$ structure.

In addition to the first and the second peaks, the third peak appears at a lower $T$ around $0.01 \lesssim T \lesssim 0.02$, although its sharpness is largely size-dependent. Furthermore, even the fourth peak appears at around the lowest $T$ studied $T\simeq 0.005$ for $N=36$, consistently with an earlier result by an approximate method~\cite{Sindzingre2}.For $N=36$, the singlet gap was estimated to be $\sim 0.010$ \cite{Waldtmann}, while the triplet gap to be much larger $\sim 0.164$ \cite{Waldtmann}. Thus, the observed low-$T$ structure of the specific heat is borne by singlet excitations. Our observation then suggests that at least the singlet gap would be quite small in the bulk, $\lesssim 0.01$, or even to be gapless.

 In the inset of the Fig.~2(b), we show the computed entropy per site $s$ in the low-$T$ range of $T\leq 0.03$ for $N=30$ and $36$. For $N=36$, with the observed fourth peak, $s$ tends to vanish in the $T\rightarrow 0$ limit observing the third law, while, for $N=30$, an additional specific-heat peak seems to be required at a still lower temperature of $T/J<0.005$.

\begin{figure}[t]
 \includegraphics[width=6.5cm, angle=-90]{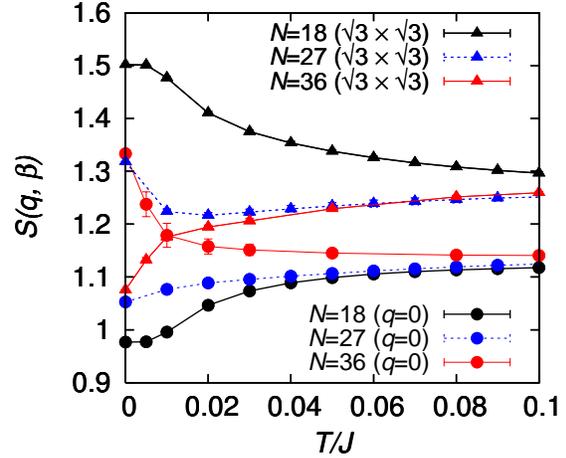}
 \caption{(Color online) The temperature dependence of the static spin structure factor $S({\bf q},\beta)$ intensities for the sizes $N=18$, 27 and 36, computed at the two $q$-points corresponding to the $q=0$ order averaged over ${\bf q}=(0,\pm 2 \pi/\sqrt{3})$ and $(\pm \pi,\pm \pi/\sqrt{3})$ [circles], and to the $\sqrt{3} \times \sqrt{3}$ order averaged over ${\bf q}=(\pm 4 \pi/3, 0)$ and $(\pm 2 \pi/3,\pm 2\pi/\sqrt{3})$ [triangles].
}
\label{}
\end{figure}

In order to get information about the nature of the spin SRO, we compute the static spin structure factor $S({\bf q},\beta)$ defined by
\begin{eqnarray}
S({\bf q},\beta)=\frac{1}{N}  \langle |\sum_{j} {\bf S}_j e^{i {\bf q} \cdot {\bf R}_j}|^2 \rangle_{\beta, N}.
\end{eqnarray}
The computed $S({\bf q},\beta)$ for $N=36$ are shown in Figs.~3 as an intensity plot in the wavevector ${\bf q}$=($q_x$,$q_y$) plane for (a) $T$=0, (b) 0.08, and (c) 0.01.

 At $T=0$, $S({\bf q},\beta)$ exhibits a broad ridge-like structure along the zone boundary of the extended Brillouin zone (BZ), in which weak SRO peaks are observed at the wavevector points corresponding to the $q=0$ state, {\it i.e.\/}, at ${\bf q}=(0,\pm 2 \pi/\sqrt{3})$ and $(\pm \pi,\pm \pi/\sqrt{3})$, the length unit here taken to be the n.n. distance of the original kagome lattice. The result is consistent with the earlier ED result by L\"auchli~\cite{Lauchli}. Note that, although the ED calculation on  smaller size clusters of $N=$18 and 27 favors the $\sqrt{3} \times \sqrt{3}$ structure than the $q=0$ structure in contrast to the present result on $N=36$ (refer also to Fig.~4 below), the recent DMRG calculation for larger systems up to $N=108$ also reported the SRO peaks appearing at the points corresponding to the $q=0$ state~\cite{Jiang}. These observations lend support to the expectation that quantum fluctuations favors the $q=0$ SRO than the $\sqrt{3} \times \sqrt{3}$ SRO in the ground state of the $S=1/2$ model.

 If one recalls the fact that the semi-classical or large-$S$ calculations suggest the preference of the $\sqrt{3} \times \sqrt{3}$ state~\cite{Chubukov, Chernyshev}, a natural expectation would be that the $\sqrt{3} \times \sqrt{3}$ SRO is favored at higher $T$ even in the $S=1/2$ model. Then, a crossover associated with the change of the dominant type of SRO might occur at a certain finite $T$ between the $q=0$ SRO at lower $T$ and the $\sqrt{3} \times \sqrt{3}$ one at higher $T$. Fig.~3(b) exhibits $S({\bf q},\beta)$ at a temperature $T$=0.08 close to the second peak of the specific heat. The dominant SRO peaks now appear at the points corresponding to the $\sqrt{3} \times \sqrt{3}$ order, {\it i.e.\/}, ${\bf q}=(\pm 4 \pi/3, 0)$ and $(\pm 2 \pi/3,\pm 2\pi/\sqrt{3})$, moving from the ones corresponding to the $q=0$ state at $T=0$ of Fig.3(a). Hence, even in the $S=1/2$ system, thermal fluctuations select the $\sqrt{3} \times \sqrt{3}$ SRO as in the classical case~\cite{Huse,Reimers,Zhitomirsky,Henley,Cepas,Chern}.

 In order to get further detailed information about the magnetic SRO, we compute the temperature dependence of the $S({\bf q},\beta)$ intensity at the two representative $q$-points corresponding to the $q=0$ and the $\sqrt{3} \times \sqrt{3}$ orders for the sizes $N=18$, 27 and 36, each  multiple of nine, and the result is shown in Fig.4. On decreasing $T$, finite-size effects get enhanced indicating the development of the magnetic SRO. At higher $T$, the $\sqrt{3} \times \sqrt{3}$ intensity exceeds the $q=0$ one irrespective of the size $N$. The behavior at lower $T$, however, turns out to differ significantly between $N=18,27$ and $N=36$. For $N=36$, on decreasing $T$, the $q=0$ intensity grows while the $\sqrt{3} \times \sqrt{3}$ intensity is suppressed, the former exceeding the latter at around $T\simeq 0.01$ close to the third-peak temperature of the specific heat as mentioned above. Thus, a finite-temperature crossover phenomenon between the two distinct types of magnetic SRO is likely to occur close to the third-peak temperature, $T \sim 0.01$.

 Of course, whether this finite-temperature crossover phenomenon survives or not in the thermodynamic limit is a nontrivial question. Yet, if one notices that the stabilization of the $\sqrt{3} \times \sqrt{3}$ SRO at higher $T$ is strongly supported both by our data of Fig.4 and by the result of the semi-classical calculations, while the stabilization of the $q=0$ SRO at lower $T$ is supported both by our data for the $N=36$ cluster and by the DMRG data of much larger sizes of $N\leq 108$~\cite{Jiang,Kolley,DMRG}, its existence is quite plausible.

 The form of $S({\bf q},\beta)$ around the crossover temperature $T_{{\rm cross}} \sim 0.01$ where the two types of SRO compete would be of special interest. In Fig.3(c), we show $S({\bf q},\beta)$ at $T=0.01$. As can be seen from the figure, the intensity here forms an almost flat ring-like ridge along the BZ boundary. It resembles the intensity of the ``ring liquid'' proposed in the frustrated honeycomb AF in Ref.\cite{Okumura} and the frustrated square ferromagnet in Ref.~\cite{Seabra}.

 We finally discuss possible implications of our present results to experiments. Recent low-energy inelastic neutron scattering measurements on $S=1/2$ kagome AF herbertsmithite revealed the broad spots corresponding to the $q=0$ SRO at a low $T$ of $T=2$K corresponding to $T \sim 0.01$~\cite{Han,Han2}. Although the authors of Refs.~\cite{Han,Han2} invoked that the impurity effects coming from the Cu$^{2+}$ impurities on adjacent triangular (Zn) interlayers as an origin of these spots, our present results indicate that the emergence of the $q=0$ SRO alone is understandable even without invoking such impurity effects. Of course, the impurity (or the quenched randomness) effect originated from the adjacent triangular layer could be important in understanding the observed spin-liquid-like behavior as recently emphasized in Refs.\cite{Kawamura}. Furthermore, the Dzyaloshinskii-Moriya interaction might be playing a role in stabilizing the $q=0$ SRO in real materials~\cite{Cepas2}. If the possible effects of the randomness and the DM interaction would be negligible in herbertsmithite (might not be the case !), the SRO peaks might move to the distinct $q$-points corresponding to the $\sqrt{3} \times \sqrt{3}$ order as the temperature is further raised.

 In summary, we studied the effects of thermal fluctuations on the ordering of the $S$=1/2 kagome Heisenberg antiferromagnet by means of the Hams-de Raedt method up to the $N=36$ cluster retaining a full symmetry of the lattice, to find that the second peak of the specific heat persists, while the third and the fourth peaks appear at lower $T$. In particular, we observed a finite-temperature crossover phenomenon occurring at $T \sim 0.01$ close to the third-peak temperature, which was associated with the changeover of the type of the magnetic SRO between the $q=0$ (lower-$T$) and the $\sqrt{3} \times \sqrt{3}$ (higher-$T$) states. 

\begin{acknowledgment}
The authors are thankful to K. Uematsu, S. Sugiura, S. Miyashita, T. Misawa, K. Yoshimi, Y. Motoyama and T. Okubo  for fruitful discussion, especially for giving us useful comments on the numerical method. This study is supported by JSPS KAKENHI Grant Number JP25247064. Our code was based on TITPACK Ver. 2 coded by H. Nishimori and the reliability was checked by means of HPhi application~\cite{HPhi}. We are thankful to ISSP, the University of Tokyo, and to YITP, Kyoto University, for providing us with CPU time.
\end{acknowledgment}


\begin{thebibliography}{50}
\bibitem{shapes} Our cluster shapes of $N=$18 and 24 are the same as $N=$18 (type2) and $N=$24 shown in Fig.~15 of Ref.~\cite{Kawamura}, respectively, while the shapes of $N=$27 and 30 are displayed in Fig.~1 of Ref.~\cite{Sugiura}.
\end{thebibliography}
\end{document}